# Waveguide-integrated black phosphorus photodetector with high responsivity and low dark current

Nathan Youngblood[1], Che Chen[1], Steven J. Koester[1], and Mo Li[1*]

[1]Department of Electrical and Computer Engineering, University of Minnesota, Minneapolis, MN 55455, USA

**Layered two-dimensional materials have shown novel optoelectronic properties and are well suited to be integrated in planar photonic circuits[1-3]. For example, graphene has been utilized for wideband photodetection[4-7]. Because graphene lacks a band gap, however, graphene photodetectors suffer from very high dark current[6,8]. In contrast, layered black phosphorous, the latest addition to the family of 2D materials[9-11], is well-suited for photodetector applications due to its narrow but finite band gap[12-17]. Here, we demonstrate a gated multilayer black phosphorus photodetector integrated on a silicon photonic waveguide operating in the near-infrared telecom band. In a significant advantage over graphene devices, black phosphorus photodetectors can operate under a bias with very low dark current and attain intrinsic responsivity up to 135 mA/W and 657 mA/W in 11.5nm and 100 nm thick devices, respectively, at room temperature. The photocurrent is dominated by the photovoltaic effect with a high response bandwidth exceeding 3 GHz.**

    Two-dimensional (2D) materials have tremendous potential for optoelectronic applications[1-3]. Graphene, the most extensively investigated 2D material, has many novel optical properties such as a tunable inter-band transition and saturable absorption, and has potential for a wide range of optoelectronic applications[1,2]. However, one important optoelectronic device application where graphene is severely limited in is photodetection. Although graphene has shown broadband optical absorption, ultrafast photoresponse and reasonable responsivity, graphene photodetectors have very high dark current when they are operated in photoconductive mode

---

[*] Corresponding author: moli@umn.edu

where a bias voltage is applied in order to attain high responsivity[6,8]. This is a direct result of the lack of a band gap in graphene. This high dark current leads to high shot noise, the dominant noise source at low levels of light, and thus sets the noise floor of the photodetector high. On the other hand, other 2D materials such as layered transition-metal dichalcogenides (TMDC) have relatively large band gaps, but typically do not absorb at telecom wavelengths[18-20].

Black phosphorus (BP) is a layered crystal of phosphorus that is stable at room temperature and has been of interest recently after experiments showing that few-layer and monolayer flakes can be exfoliated, similar to graphene[9-11]. In contrast to graphene, however, BP has a direct band gap predicted to be 1.8-2.0 eV for monolayer and as the number of layers increases, the band gap reduces and eventually reaches ~0.3 eV in bulk[21,22]. Such a layer tunable band gap covering the visible to the mid-infrared spectral range suggests that BP is a very promising 2D material for broadband optoelectronic applications[13,14]. Additionally, BP has shown excellent electrical properties including a high hole mobility up to 1000 $cm^2V^{-1}s^{-1}$ at room temperature, an on-off ratio up to $10^5$ and good current saturation in field-effect devices[9-11]. In this work, we integrate a few-layer BP photodetector in a silicon photonic circuit, which also enables quantitative measurement of the absorption and the quantum efficiency of BP at the important telecom band (~1.55 μm). Operated in photoconductive mode, high responsivity, high internal quantum efficiency and operation at bitrates above 3 Gbit/s are achieved with very low dark current.

To most efficiently utilize 2D materials' optoelectronic properties, it is necessary to integrate them on planar photonic devices so that the optical interaction length is not bounded by the thickness of the materials[6,7,23-28]. Here, we integrate on a silicon photonic waveguide a BP field-effect transistor (FET) using few-layer graphene as the top-gate. In this way, the carrier type and concentration in the BP layer can be electrostatically tuned, allowing the mechanism of photocurrent generation to be investigated and the performance of the device to be optimized. Fig. 1a illustrates the anatomy of the device, featuring the BP channel and the graphene top gate. Fig. 1b shows an optical image of the completed device. The silicon photonic circuit patterned on the silicon-on-insulator (SOI) substrate includes a Mach-Zehnder interferometer (MZI). This allows the absorption in the BP to be quantitatively measured in order to unequivocally determine the intrinsic responsivity and internal quantum efficiency[23,26], which are more useful than the extrinsic

-2-

values to reveal BP's true potential for optoelectronic applications. The BP layer was exfoliated from a bulk crystal and transferred onto one of the optical arms (one measurement arm, one reference arm) of the MZI using a wet transfer method[29], after planarization of the substrate with silicon dioxide. Fig. 1c shows an atomic force microscope image of the BP layer transferred onto the waveguide after deposition of source and drain contacts made of titanium (5nm) and gold (50 nm). The width of the BP is 6.5 μm and the thickness is determined to be 11.5 nm from the profile shown in Fig. 1c, indicating it has about 23 layers. During each step of fabrication (Fig. 1d), the transmission spectrum of the MZI was measured. From the extinction ratio (*ER*) measured from the interference fringes in the spectra, the absorption loss in the measurement arm of the MZI can be determined using the relation: $ER = \left(1+e^{\alpha \Delta L/2}\right)^2 / \left(1-e^{\alpha \Delta L/2}\right)^2$, where $\alpha$ is the absorption coefficient and $\Delta L$ is the width of the BP FET. The results are shown in Fig. 1d. From the measurement after the transfer of BP and the deposition of the top gate dielectric (see Methods), we determine that the absorption coefficient in BP in the device is 0.182 dB/μm. This result confirms that few layer BP has strong absorption in the near-infrared as expected from its narrow band gap[13]. The 6.5 μm long device absorbs 78.7% of the optical power in the waveguide, of which 17.5% is absorbed by the BP and the rest by the graphene top-gate, which is the main cause of loss. However, such a top-gate is unnecessary in practical devices if the initial doping in the BP can be optimized by, for example, chemical doping.

We first performed field-effect transport measurements of the device to characterize its electrical properties. The source-drain current ($I_{DS}$) under sweeping the gate voltage ($V_G$) at various fixed bias voltages ($V_{DS}$) is shown in Fig. 2a. The ambipolar transport behavior of the BP is clearly shown. The minimum conductance point at $V_G$=-7 V reveals that the BP is n-doped with an electron concentration estimated to be $1.6 \times 10^{13}$ cm$^{-2}$. We have consistently found that in devices using top gates with an aluminum oxide ($Al_2O_3$) gate dielectric grown with atomic layer deposition (ALD), the BP is n-doped. This is in contrast to p-doped BP observed in most of the back gated devices reported in the literature so far[9-11] and in devices that we have made using hafnium oxide ($HfO_2$) as the top gate dielectric. Therefore, we speculate that the doping type and level in BP is sensitive to the type of the dielectric layers above and below the BP and to the deposition conditions of those layers. This relationship will require more systematic studies to draw meaningful conclusions but is not the focus of this work. With the configuration of a top-gate and thin gate dielectric, the source-drain current of our device can be efficiently modulated by the gate voltage with an on-off ratio of 500, similar to other devices with comparable BP thicknesses[10,11,22]. However, unlike BP devices using a bottom-gate[9-11], the field effect characteristics (Fig. 2a) of our top-gated BP FET shows much higher conductance on the electron side, when a large positive gate voltage is applied, than on the hole side. Also, the source-drain *I-V* relation is linear when the BP is n-doped but nonlinear when the BP is p-doped as shown in Fig. 2b and c. This indicates that the Schottky barrier at the contact-BP interface is lower when the BP is n-doped than when the BP is p-doped. This can be attributed to the effects induced by capping the BP with the $Al_2O_3$ gate dielectric layer[30]. Additionally, the fact that the *I-V* characteristics are mostly symmetric about the positive and negative source-drain voltage reflects that the two contacts are nearly identical. Therefore, excellent transport property and efficient tuning of the carrier type and concentration in BP are achieved in our device.

We next measured the BP device's response to optical signals in the waveguide. A tunable diode laser was fiber coupled to the grating couplers, which have a coupling efficiency of ~25%. Fig. 3a shows the *I-V* characteristics of the device at various optical power levels and with $V_G$=-8 V so the BP is nearly intrinsic. It can be clearly seen that significant photocurrent, positive relative to the $V_{DS}$, is generated when the optical signal is on. The positive sign of photocurrent suggests that the generation mechanism is photovoltaic[8,12,15]. The photo-thermoelectric effect can be ruled out as the photogeneration mechanism because the device is symmetric between the source and the drain contacts[8,12,31]. However, when $V_G$ is increased to be more positive so the BP is gated to be more heavily n-doped, as shown in Fig. 3b, the sign of photocurrent changes to be negative relative to the $V_{DS}$. The sign change suggests that the photocurrent generation mechanism is no longer photovoltaic, but due to the bolometric effect which stems from the decreased conductance when the BP is heated by optical absorption. This bolometric effect is similar to that observed in biased graphene photodetectors, in which the reduction of conductance at elevated temperature is attributed to increased phonon scattering[8,12]. Compared with graphene, it appears that the bolometric photocurrent in BP at high doping is much weaker than the photovoltaic current at low doping, which is attributed to the higher quantum efficiency in BP than in graphene. To further illustrate these two distinct regions of photogeneration, we systematically measured the photocurrent at various gate and bias voltages with a fixed optical power of 1.91 mW. The results are plotted in Fig. 3c. The change of the photocurrent sign with the gate voltage is clearly seen for all bias voltages. At low doping (-10 V<$V_G$<-1 V), the photocurrent is strong and dominated by the photovoltaic current; whereas at high n-type doping ($V_G$>0 V), the photocurrent is weak, opposite the bias voltage and dominated by the bolometric effect. We calculate the intrinsic responsivity and internal quantum efficiency using the actual amount of optical power absorbed by BP determined from the interferometry results. As can be seen from Fig. 3d, optimal performance is attained when the BP is gated to be nearly intrinsic when $V_G$=-8 V and under a large bias voltage $V_{bias}$=$V_{DS}$=-0.4 V, achieving the best intrinsic responsivity of 135 mA/W and internal quantum efficiency of 10%. However, taking the absorption by the graphene top-gate into account, the overall extrinsic responsivity of the device is reduced to 18.8 mA/W. Higher responsivity could be achieved with larger bias voltage. But because the applied gate voltage is high, the bias voltage was not increased beyond ±0.4 V to avoid the risk of breaking-down the gate dielectric. In another device with 100 nm thick BP and no top-gate, as shown in Fig. 3e, intrinsic responsivity up to 657 mA/W and internal quantum efficiency up to 50% are achieved at +2 V bias voltage (see Supplementary Information).

Importantly, due to the finite band gap in BP, the dark current of our device is very low. At the optimal operation condition ($V_G$=-8 V, $V_{bias}$= -0.4 V), the dark current $I_{dark}$ is only 220 nA, which is more than three orders of magnitude less than that of graphene photodetector under similar bias voltage[5,8]. Taking the external responsivity into account, the normalized photocurrent to dark current ratio (NPDR)[32] of our BP photodetector is plotted in the inset of Fig. 3a. At $V_{bias}$=-0.4 V, the NPDR of our BP photodetector is 85 mW$^{-1}$, which is four orders of magnitude higher than that of typical graphene devices[5,6]. This value of NPDR is also comparable to that of waveguide-integrated germanium photodetectors of similar configuration[33] and could be improved by an order of magnitude if the top gate absorption loss was eliminated. Even though graphene photodetectors can be operated in photovoltaic mode at zero bias, with zero dark current but with a substantially compromised responsivity[5,6], the low resistance of graphene leads to high Johnson current noise from the shunt resistance. In comparison, the source-drain shunt resistance of our BP photodetector is very high ($R_{SD}$>1 MΩ) so the noise of the photodetector is dominated by dark current shot noise. Thus, in terms of both responsivity and noise level, the BP photodetector has significant advantages over its graphene counterparts.

The frequency response of the photocurrent can also help reveal the mechanism of photocurrent generation. Fig. 4a shows the normalized response of the photodetector to an amplitude modulated optical signal over a broad frequency range from 10 Hz to 10 GHz, measured by using a lock-in amplifier (for less than 100 kHz) and network analyzers (for 100 kHz to 10 GHz) while fixing the bias voltage ($V_{bias}$=0.4 V). When the BP is gated with $V_G$=-8V to be nearly intrinsic, the photodetector shows a high-speed response with a roll-off frequency measured to be $f_{-3dB}$=2.8 GHz, which is limited by the RC bandwidth of the device as well as the bandwidth of measurement instruments including the current amplifier and the bias tee. This fast response is expected for the photovoltaic mechanism of photocurrent generation which is fundamentally limited by the carrier recombination time in BP and is accelerated by the bias field. In a stark contrast, when the BP is gated to be highly n-doped, the photocurrent response rolls-off at a much lower frequency of $f_{-3dB}$=0.2 MHz. Because the photocurrent is in this regime is due to the bolometric effect, this slower frequency response can be attributed to the low in-plane lattice thermal conductivity of BP[34], as well as the thermal response of the silicon waveguide[35]. The dramatically different response speed at different doping levels in the BP thus further corroborates the distinct generation mechanisms and highlights the optimized performance of the device when BP is gated to be intrinsic. To demonstrate the feasibility of using the BP device in high-speed optical communication, we also performed an eye-diagram measurement using a pseudo-random bit series (PRBS) with a data rate of 3 Gbit/s and a low optical power level of 1.2 mW (0.8 dBm). The completely open eye shown in Fig. 4b indicates the BP photodetector can be readily used for practical optical communication.

In conclusion, a waveguide integrated and gate tunable photodetector based on few-layer black phosphorus for the telecom band has been demonstrated. High responsivity, high response speed and low dark current are achieved when the BP is gated to low doping and the photocurrent generation is dominated by the photovoltaic effect. In nearly every aspect of performance, the BP photodetectors can outperform graphene photodetectors and are more ready for practical use. We expect the performance of BP photodetectors to be further improved when larger flakes of BP can be exfoliated or large scale growth of BP with high quality can be achieved. If the number of layers in BP can be controlled, its band gap can be tailored for a specific wavelength so that both responsivity and dark current of the photodetector can be optimal. In addition, BP shows strong anisotropy in both dc and optical conductivity[10,36]. If the armchair direction of BP can be aligned with the source-drain direction (zigzag along the waveguide) using Raman spectroscopy (see SI) so that both optical absorption and carrier mobility are maximal, the responsivity and the speed of the photodetector can be maximized with a reduced device footprint. These results and prospects of black phosphorus, combined with prior demonstrations of many optoelectronic devices based on graphene and transition-metal dichalcogenides, indicate that 2D materials and their heterostructures[17,37-39] can provide the necessary components for realizing complete optical communications links and chart a clear path toward the commercial viability of 2D materials for integrated optoelectronic applications.

**Methods:**

Black phosphorus photodetectors were fabricated using silicon-on-insulator (SOI) wafers (SOITEC Corp.) with 110 nm top silicon layer and 3 μm buried oxide layer. The underlying photonics layer was patterned using electron beam lithography (Vistec EBPG 5000+) using maN-2403 resist and etched with a standard silicon Bosch process to define the photonics layer. Electron beam evaporation was then used to deposit 140 nm silicon oxide on the sample using the remaining ebeam resist as a mask. After removing the resist in NMP with an ultrasonic bath, the planarized substrate was annealed using rapid thermal annealing (RTA) at 1100 ºC for 1 minute to improve the quality of the evaporated oxide. A 10 nm layer of $HfO_2$ was grown with ALD to protect the photonics layer from subsequent etching processes. Exfoliated black phosphorus (purchased from Smart Elements GmbH) was transferred to the planarized photonics layer according to the wet transfer method described in Ref. 29. Careful consideration was given to prevent the black phosphorus from making contact with the water. A photolithography and dry etch step was used to remove unwanted material from the waveguides. Shipley S1800 series photoresist was used to protect the black phosphorus channel material during the dry etch. Source and drain contacts were defined with ebeam lithography using PMMA and 5 nm Ti / 50 nm Au was deposited using ebeam evaporation at 100 ºC. A 20 nm ALD $Al_2O_3$ gate dielectric / passivation layer was then grown at 250 ºC. The top few-layer graphene gate was transferred using the same wet transfer method and patterned with ebeam lithography and oxygen plasma etching. Finally, top gate contacts were patterned using PMMA and 5 nm Ti / 30 nm Au was deposited with ebeam evaporation.

Photocurrent maps at different biases and gate voltages were measured at a fixed wavelength for various laser powers. Two Keithley 2400 series source meters were used in this measurement, one to control the gate voltage and the other to bias the device while measuring the source-drain current. The devices were held at a fixed gate while the bias voltage was swept twice – once with no optical power and once with the laser source on. The difference between the two scans yielded gate and bias dependent photocurrent. Power and bias dependent photocurrent was extracted from these two dimensional scans and responsivity was calculated after accounting for loss due to the input grating coupler, power splitting arm of the MZI structure, and power absorbed by the channel material.

Low frequency photoresponse (10 Hz to 100 kHz) was measured by modulating the laser source with a high bandwidth electro-optic modulator (Lucent 2623NA) driven by a signal generator. Photocurrent was amplified using a low-noise current preamplifier (Stanford Research Systems, SR570) and monitored with a Stanford Research Systems, SR830 lock-in amplifier. The gate voltage was swept from -10 V to +10 V and the amplitude of the photocurrent was recorded at various driving frequencies. Mid-range frequencies (100 kHz to 10 MHz) were measured using a RF pre-amplifier and a 3 GHz bandwidth network analyzer (HP 3577B). High frequency measurements (10 MHz to 10 GHz) were achieved with a 20 GHz bandwidth network analyzer (Agilent E8362B PNA). A 12 GHz photoreceiver (Newport 1554-A) was used to monitor the frequency dependence of the EOM and correct the frequency response measured. All measurements were performed at room temperature in atmosphere.

**Acknowledgements**

This work is supported by the National Science Foundation (Award No. ECCS-1351002) and the Air Force Office of Scientific Research (Award No. FA9550-14-1-0277). Parts of this work was carried out in the University of Minnesota Nanofabrication Center which receives partial support from NSF through NNIN program, and the Characterization Facility which is a member of the NSF-funded Materials Research Facilities Network via the MRSEC program.

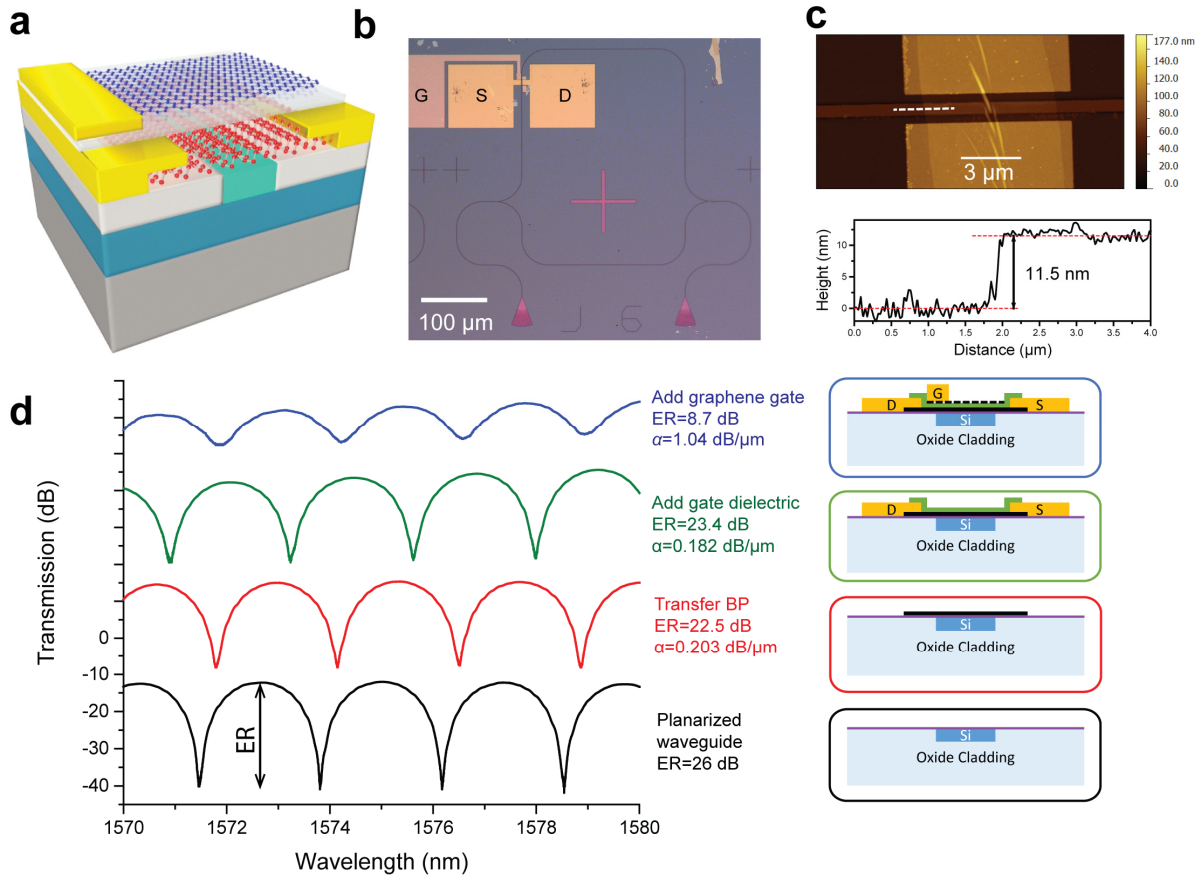

**Figure 1 Black phosphorus photodetector integrated in silicon photonic circuit. a)** 3D illustration of the device configuration, featuring a few-layer graphene top-gate. **b)** Optical microscope image of a completed device. A BP field-effect transistor is integrated in one arm of a Mach-Zehnder interferometer circuit. **c)** Atomic force microscope image of the BP with contacts and before the deposition of gate dielectrics and the fabrication of the top-gate. The bottom inset is the height profile along the white dashed line in the top panel, showing the thickness of the BP is 11.5 nm, corresponding to 23 monolayers. **d)** Transmission spectra of the Mach-Zehnder interferometer measured after each step of fabrication as illustrated in the schematics listed in the right column. From the extinction ratio, the absorption coefficient of each added layer can be determined. With this method, the absorption coefficient in the BP in the completed device is determined to be 0.182 dB/µm.

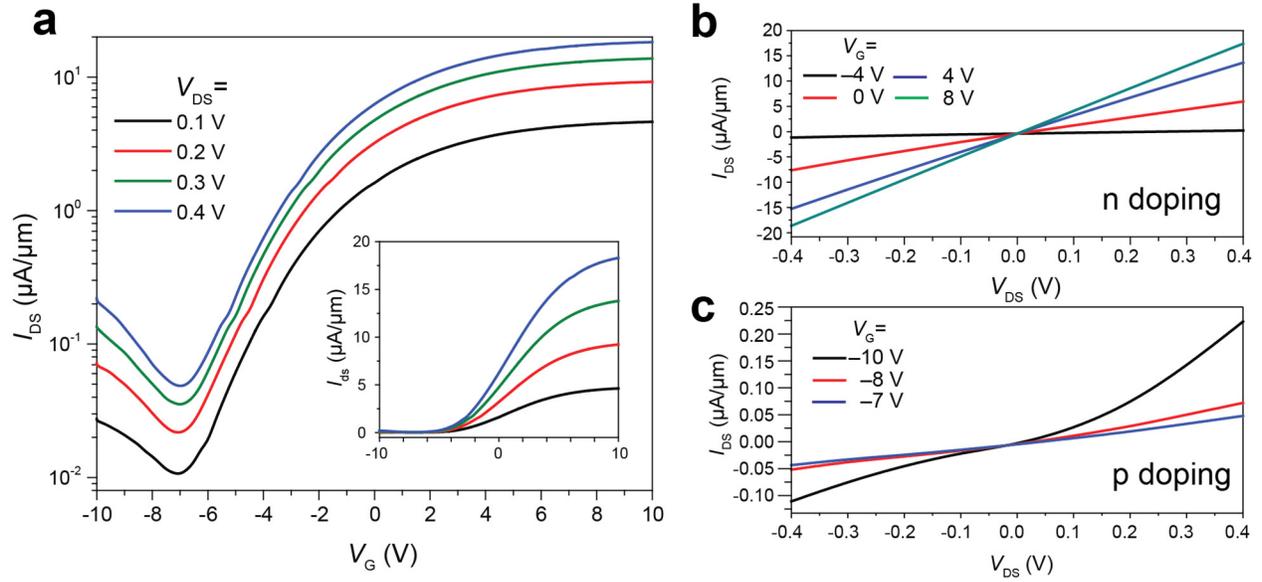

**Figure 2 Field effect characteristics of the BP photodetector. a)** Source-drain current ($I_{DS}$) of the device at various fixed bias voltages ($V_{DS}$) as the gate voltage ($V_G$) is swept. Inset: linear scale plot of the main panel. **b)** and **c)** *I-V* characteristics of the device when the BP is gated to be n-doped (panel b) and p-doped (panel c), respectively.

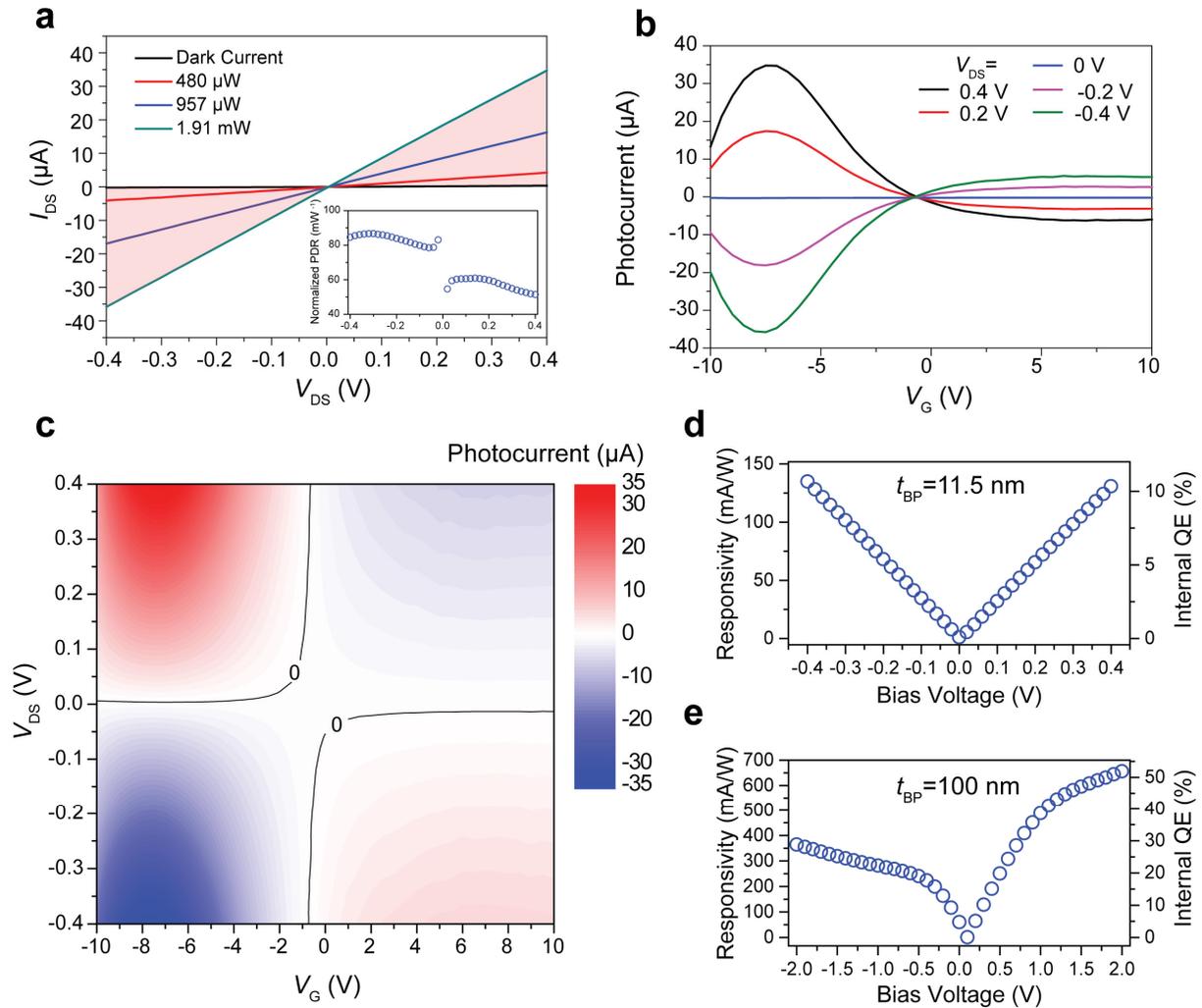

**Figure 3 Gate and bias tuned photoresponse of the BP photodetector. a)** Source-drain current versus bias voltage when optical signal is off (black line) and at various power levels (red: 480 μW; blue: 957 μW; green: 1.91 mW). The gate voltage is fixed at -8 V. The optical power values refer to the power in waveguide underneath the BP. The pink shadowed area indicates the contribution of photocurrent. Inset: Normalized photocurrent to dark-current ratio (NPDR) for an optical power of 1.91 mW. It is noteworthy that the dark current of the device under a bias voltage is very low. **b)** Photocurrent versus gate voltage at various fixed bias voltages. The direction of the photocurrent changes as the gate voltage increases from negative values (where the BP has low doping) to positive values (where the BP is highly n-doped). The optical power in the waveguide is fixed at 1.91 mW. **c)** 2D contour plot of the photocurrent as a function of both gate and bias voltages. In the low doping region, the photocurrent generation is dominated by photovoltaic

effect, whereas in the highly n-doped region, the bolometric effect dominates the photoresponse. **d)** Intrinsic responsivity and internal quantum efficiency of the BP photodetector determined using the measured photocurrent and the optical power that is absorbed by the BP. A responsivity of 135 mA/W and internal quantum efficiency of 10% are achieved with bias voltage of -0.4 V. **e)** Intrinsic responsivity and internal quantum efficiency of another device with 101 nm thick BP and under larger bias voltage. A high responsivity of 657 mA/W and internal quantum efficiency of 50% are achieved with bias voltage of 2 V.

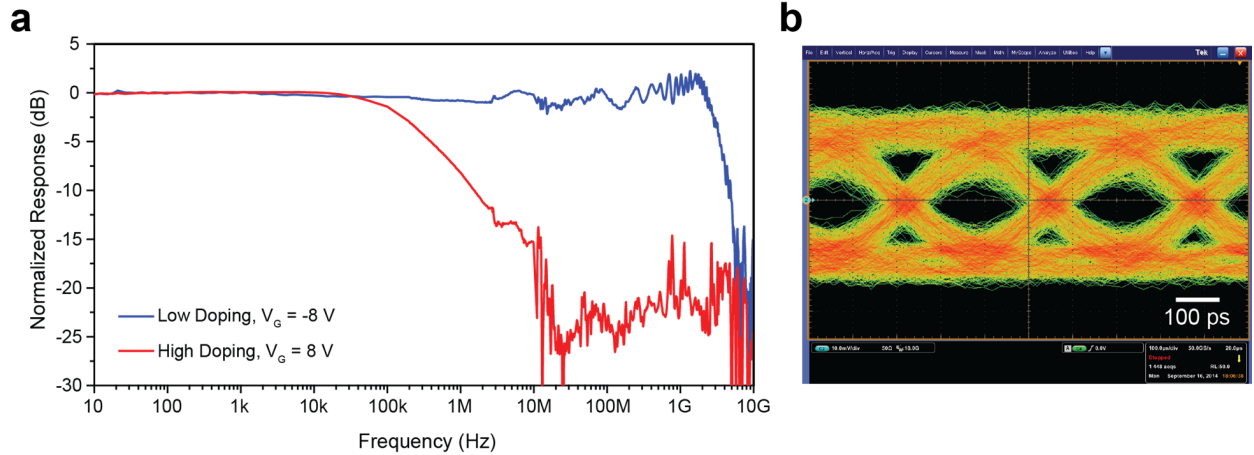

**Figure 4 Broadband frequency response of the BP photodetector. a)** The response of the BP photodetector is measured when the BP is gated to low and high doping. At low doping, the response is broadband with a cut-off frequency of 3 GHz, which is limited by the RC bandwidth of the contact pads and the input impedance of the preamplifier. At high doping, the response rolls-off at 0.2 MHz, indicating that the photoresponse is of thermal origin as expected from the bolometric effect. **b)** Receiver eye diagram at 3 Gbit/sec data rate measured with the BP photodetector. Scale bar: 100 ps.